\documentstyle[11pt,aaspp4]{article}

\begin{document}

\title{A Sub-Relativistic Shock Model for the Radio Emission of
SN1998bw}
\author{Eli Waxman$^1$ and Abraham Loeb$^2$}
\medskip
\noindent
1. Department of Condensed-Matter Physics, Weizmann Institute, Rehovot
   76100, Israel\break
\smallskip
\noindent
2. Astronomy Department, Harvard University, 60 Garden Street, Cambridge,
MA 02138, USA

%\altaffiltext{2}{email:aloeb@cfa.harvard.edu}

\begin{abstract}

SN1998bw is the most luminous radio supernova ever observed.  Previous
discussions argued that its exceptional radio luminosity, $\sim 4\times
10^{38}~{\rm erg~s^{-1}}$, must originate from a highly relativistic shock
which is fully decoupled from the supernova ejecta. Here we present an
alternative model in which the radio emission originates from a
sub-relativistic shock with a velocity $\simeq 0.3c$, generated in the
surrounding gas by the expanding ejecta. In this model, thermal electrons
heated by the shock to a relativistic temperature of $\sim 60$ MeV, emit
synchrotron self-absorbed radiation in the post-shock magnetic field. This
model provides an excellent fit to the observed spectra provided that the
thermal electrons are in equipartition with the ions behind the shock. The
required magnetic field is much weaker than its equipartition value and
could have been carried out by the progenitor's wind prior to the supernova
explosion. According to this model, the radio emission from SN1998bw is
unrelated to the highly relativistic blast wave that produced the
$\gamma$-ray burst GRB980425.

\end{abstract} 

\keywords{supernovae: individual (SN1998bw) -- radio continuum: general --
gamma rays: bursts}

\section{Introduction}

The optical emission spectrum of SN1998bw classifies this event as a Type
Ib/Ic supernova (Lidman et al. 1998; Sadler et al. 1998; Patat \& Piemonte
1998; Galama et al. 1998), suggesting that it resulted from a core collapse
of a massive star which lost its hydrogen and helium envelope. Radio
monitoring of the source revealed an exceptionally bright emission which
peaked after $12\pm 2$ days (Kulkarni et al. 1998).  The inferred
radio luminosity at the source redshift of 0.0083, $\sim 4\times
10^{38}~{\rm erg~s^{-1}}$, is the highest ever observed for a
supernova. Multi-frequency measurements around the peak of the lightcurve
showed a strong suppression of the emitted flux below a frequency of a few
GHz.

Aside from being associated with a rare class of supernovae, SN1998bw also
occurred inside the 8 arcminute error circle of the Gamma-Ray Burst
GRB980425 which was detected by the Beppo-SAX satellite at about the same
time (Soffitta et al. 1998). The small likelihood for a chance overlap
between the two events, $\sim 10^{-4}$, led to the suggestion that the two
might be associated (Galama et al. 1998) and to the conjecture that perhaps
all SN Ib,c events lead to Gamma-Ray Bursts (Wang \& Wheeler 1998; Woosley,
Eastman, \& Schmidt 1998).  However, the non-Euclidean number-count
statistics, the event rate, and the temporal and spectral properties of
most GRBs imply that SN1998bw-like events could only be associated with a
minority fraction, $\la 10\%$ of the GRB population (Bloom et al. 1998).
Indeed, the lack of an X-ray afterglow at the position of SN1998bw makes
its potential GRB counterpart rare, since all but one of the other
11 GRBs detected so far by Beppo-SAX 
showed an X-ray afterglow. In fact, a separate X-ray transient source,
1SAXJ1935.3-5252, was identified inside the Beppo-SAX error circle, with a
flux comparable to that of the X-ray afterglows of other GRBs.
This source provides an alternative to the SN-GRB association.  A chance
overlap between the error circle of GRB980425 and SN1998bw would fit better
theoretical expectations, since supernova models have difficulties
accounting for the highly-relativistic shock required to produce a GRB
(Woosley et al. 1998). Hence, in gauging the likelihood of an association
between SN1998bw and GRB980425 it is of fundamental importance to
understand whether the bright radio emission from SN1998bw requires a
relativistic shock by itself. If it does, then a more relativistic
incarnation of the same shock at earlier times would serve as a plausible
source for GRB980425 and hence strengthen the case for an association
between the two events.

The velocity of the supernova ejecta was measured from the blue wing of the
Ca II line to be $\sim$ 60,000 ${\rm km~s^{-1}}=0.2 c$ about a week after the
explosion (Stathakis et al. 1998; Kulkarni et al. 1998).
Kulkarni et al. (1998) argued that the radio emission must have originated
from a shock which was fully decoupled from the main supernova ejecta and
expanded at a relativistic speed. They based their assertion on two
arguments:

%\itemize{

\noindent
1. If the radio source expanded at the measured ejecta velocity, then the
brightness temperature of the source exceeded the threshold temperature for
the so-called ``inverse Compton catastrophe'' (Kellerman \& Pauliny-Toth
1969, Readhead 1994). Under these conditions, Compton scattering of the
radio photons to higher energies would have dominated the luminosity and
violated upper limits on the X-ray flux of SN1998bw (Pian et
al. 1998a,b). Relativistic expansion alleviates this problem.

\noindent
2. The lack of a strong variability at low frequencies implies that the
source size exceeded the refractive scintillation scale of $\sim
10^{16}~{\rm cm}$ at the peak of the lightcurve, and therefore expanded at
a speed $\ga 0.3c$.

In order to put these constraints in a concrete physical context, we have
constructed the simplest model for synchrotron emission due to the
sub-relativistic ejecta of SN1998bw.  In this model, the ejecta push as a
piston and generate a strong forward shock in the surrounding gas (the
latter being a remnant of the progenitor's wind prior to the supernova
explosion). The shock front heats electrons to a relativistic temperature,
and these emit incoherent synchrotron radiation in the post-shock magnetic
field. To our surprise, we had found this model to be consistent with all
observational data. In particular, the two constraints mentioned above are
satisfied for the following reasons:

\noindent
1. The relativistic electrons responsible for the observed synchrotron
radiation have a typical Lorentz factor of only $\sim 360$, and hence
scatter the radio photons up in energy only into the optical regime but not
into the X-ray band.  Hence, the upper limit set by Beppo-SAX on the X-ray
luminosity of SN1998bw is not in conflict with this model.  In fact, the
X-ray luminosity is dominated by scattering of the optical supernova
photons by the relativistic electrons. This flux is consistent with the
BeppoSAX bound for an expansion speed $\ga 0.3c$ or a corresponding bulk
Lorentz factor of only $\ga 1.05$.

\noindent
2. We find that a shock velocity of $\simeq 0.3c$ is sufficient for
explaining the radio properties of SN1998bw.  This velocity is $\sim 50\%$
higher than the measured velocity shift of the CaII line. However, since
the energy ($\sim 10^{49}$~erg) and mass ($\sim 10^{-4}M_\odot$) associated
with the radio emitting plasma are small fractions of the total energy and
mass of the ejecta, the required shock can be produced by a small amount of
material preceding the bulk of the ejecta.  In fact, our results are
consistent with the modeling of the optical emission from SN1998bw by
Woosley et al. (1998, see their \S 3.2), who suggested that the outer
$10^{-3}M_\odot$ of the supernova ejecta might have moved at a speed of
$\sim c/3$ and carried $\sim 10^{50}$~ergs.

In section \S2 we describe our model and derive the expected emission
spectrum. In \S3 we compare model predictions with the radio data, and
derive the physical properties of the emitting electrons. The implications
of our results are discussed in \S4. 
Throughout the discussion, we adpot a distance of $d=38$ Mpc to SN1998bw,
assuming that it is located in the galaxy ESO 184-G82 at a redshift
$z=0.0083$ (Lidman et al. 1998; Tinney et al. 1998) and that the Hubble
constant is 65 ${\rm km~s^{-1}~Mpc^{-1}}$.

\section{Model}

In our model, the radio luminosity of SN1998bw is emitted by the material
behind a forward shock, which is generated by the expanding ejecta in the
medium surrounding the supernova progenitor.  Since the ejecta expand at a
sub-relativistic velocity, we assume to leading order that the radio flux
observed at any given time is emitted by a static spherical shell of radius
$r$. The compressed material behind the shock is expected to occupy a thin
shell of width $r/\eta$, with
%\footnote{For a strong shock with a compression ratio of 4 and a 
%uniform shell, mass conservation implies
%$\eta=12$.}  
$\eta\approx 10$. We denote the mean number density of
electrons inside this shell by $n$.

The energy distribution of radiating electrons can be constrained from the
observed synchrotron spectrum of the source.  Kulkarni et al. (1998) have
assumed that the radiating electrons have a power-law energy
distribution. The synchrotron self-absorption spectrum of such an electron
population is expected to show a spectral flux index of $2.5$ at low photon
frequencies (Rybicki \& Lightman 1979). However, as noted by Kulkarni et
al. (1998), the observed spectral index is closer to $\sim 2$. This
indicates that the electron distribution is truncated at low energies, as
would be the case if they were thermal.  For simplicity, we assume a
mono-energetic electron population which should mimic closely the emission
properties of thermal electrons. As shown below (see Figs. 1 and 2), our
simple model provides an excellent fit to the observed spectra.

The synchrotron power per unit frequency $\nu$ emitted by a single electron
of Lorentz factor $\gamma$ is given by (Rybicki \& Lightman 1979),
\begin{equation}
P(\nu,\gamma)={e^3 B\over m_e c^2} F\left[{\nu \over
\nu_c(B,\gamma)}\right],
\label{eq:single_e}
\end{equation}
where $e$ and $m_e$ are the electron charge and mass, $B$ is
the post-shock magnetic field strength and
\begin{equation}
\nu_c\equiv \gamma^2\left({eB\over 2\pi m_ec}\right).
\label{eq:frequency}
\end{equation}
The function $F(x)$ describes the synchrotron power spectrum (Rybicki \&
Lightman 1979), averaged over an isotropic distribution of pitch angles.
The total number of electrons in the radiating shell is $n\times (4\pi
r^3/\eta)$, and so the total flux per unit frequency observed at a distance
$d=38~{\rm Mpc}$ is,
\begin{equation}
f_\nu\approx \left({1-e^{-\tau_\nu}\over \tau_\nu}\right) 
\left({n r^3 \over \eta d^2}\right) P(\nu,\gamma),
\label{eq:flux}
\end{equation}
where $\tau_\nu=\alpha_\nu\times (r/\eta)$ is the optical depth per unit
frequency for synchrotron self-absorption across the shell thickness.
The self-absorption coefficient is given by (Rybicki \& Lightman 1979),
\begin{equation}
\alpha_\nu=-{1\over 8\pi m_e\nu^2}\int d\gamma^\prime 
P(\nu,\gamma^\prime)\gamma^{\prime 2}
{\partial \over \partial\gamma^\prime}
\left( {1\over \gamma^{\prime 2}}{dn\over d\gamma^\prime}\right).
\label{eq:optical_depth}
\end{equation}
For mono-energetic electrons of Lorentz factor $\gamma$,
$dn/d\gamma^\prime=n\delta(\gamma^\prime-\gamma)$, and the absorption
coefficient can be integrated by parts. We then get,
\begin{equation}
f_\nu= A\nu^2 \xi\left({\nu\over \nu_c}\right),
\label{eq:totalflux}
\end{equation}
where the functions,
\begin{equation}
\xi(x)\equiv {1-e^{-\tau_\nu}\over 1- [d\ln F(x)/d\ln x]},
\label{eq:xi}
\end{equation}
\begin{equation}
\tau_\nu(x)\equiv 
\tau_c x^{-2}F(x) \left[1- {d\ln F(x)\over d\ln x}\right],
\label{eq:tau}
\end{equation}
and the constants,
\begin{equation}
A\equiv 4\pi \gamma m_e\left({r\over d}\right)^2,
\label{eq:A}
\end{equation}
\begin{equation}
\tau_c\equiv {1\over 8\pi \nu_c^2}\left({2 n r\over \eta \gamma m_e}\right)
\left({e^3B\over m_ec^2}\right).
\label{eq:tau_c}
\end{equation}
The shape of the model spectrum, $\xi(\nu/\nu_c)$, is determined by a
single dimensionless parameter, $\tau_c$. The frequency and flux
normalizations are determined by the dimensional parameters $\nu_c$ and
$A$.

\section{Comparison with Observations}

The most stringent constraints on the model parameters are obtained, as
shown below, from observations near the peak of the radio lightcurve,
$\sim12$ days after GRB980425. These constraints are derived in
\S3.1. Observations at later times are discussed in \S3.2.

\subsection{Observations Near the Peak of the Radio Lightcurve}

Figure 1 shows the observational data from the first two epochs (12 and 15
days after GRB980425) when multi-frequency measurements were taken around
the peak of the radio lightcurve (Kulkarni et al. 1998). Our simple model
of emission from mono-energetic electrons (solid line) provides an
excellent fit to the data. The spectral shapes at both epochs are
consistent with the model spectral shape in equation~(\ref{eq:xi}) for
$\tau_c=0.6$. The frequency and flux normalizations are $\nu_c=7.6$ GHz,
$A=3.3\times10^{-44}$ g at the first epoch, and $\nu_c=6.2$ GHz,
$A=4.0\times10^{-44}$ g at the second epoch.  For comparison, we also show
the predicted spectra for a power-law population of electrons,
$dn/d\gamma\propto \gamma^{-p}$, with $p=2$ (dashed line) or $p=3$
(dotted), which are not consistent with the data. Note that the shape of
the low frequency tail is independent of all parameters in both models.

The free variables in our model are the density and temperature (Lorentz
factor) of the radiating electrons, and the magnetic field strength behind
the shock.  The spectral constraints 12 days after GRB980425 can be
parametrized in terms of the radius of the emitting shell,
$r_{16}=(r/10^{16}~{\rm cm})$, the synchrotron frequency,
$\nu_8\equiv(\nu_c/8)$ GHz, the flux amplitude, $A_3\equiv (A/3\times
10^{-44}~{\rm g}$), and the optical depth coefficient $\tau_{.6}\equiv
(\tau_c/0.6)$. We adopt a value of $\eta=10$, and note that $r_{16}=1$
corresponds to an average expansion speed of $\simeq 0.3 c$.

Equation~(\ref{eq:A}) yields the electron Lorentz factor,
\begin{equation}
\gamma\approx 3.6\times 10^2 
 A_3 r_{16}^{-2} ,
\label{eq:Lorentz}
\end{equation}
which is equivalent to an electron temperature of $kT\equiv {1\over 3}
\langle \gamma m_e v^2\rangle \approx 60 A_3 r_{16}^{-2}$ MeV.  The
magnetic field strength can then be obtained from
equation~(\ref{eq:frequency}),
\begin{equation}
B\approx 2.2\times 10^{-2} \nu_8 A_3^{-2} r_{16}^4~{\rm G}.
\label{eq:magnetic_field}
\end{equation}
Substitution of these results into equation~(\ref{eq:tau_c})
yields,
\begin{equation}
n\approx 5.3\times 10^4\nu_8 \tau_{.6} A_3^3 r_{16}^{-7}~{\rm cm^{-3}} .
\label{eq:density}
\end{equation}

The inferred number density of radiating electrons is much higher than that
of the interstellar medium and is likely to be associated with a wind that
originated from the supernova progenitor before or during the
explosion. The existence of a dense ambient medium is natural given that
the progenitor might have lost its hydrogen and helium envelope prior to
the explosion (Patat \& Piemonte 1998).

The mass of ions associated with the radiating electrons is small, 
\begin{equation}
M_{\rm ion}\approx 1.1\times 10^{-4} \mu_2 \nu_8 \tau_{.6}A_3^3 
 r_{16}^{-4}~M_\odot,
\label{eq:M}
\end{equation}
where $\mu=2\mu_2$ is the atomic mass per free electron.
For fully-ionized metal-rich material, $\mu_2 \sim 1$. The
total shock energy associated with this mass at a post-shock fluid speed
$V=0.3 V_{.3} \times c$ is given by
\begin{equation}
E_{\rm shock}\approx M_{\rm ion} V^2= 1.7\times 10^{49} \mu_2 V_{.3}^2
\nu_8 \tau_{.6}A_3^3 r_{16}^{-4}~{\rm ergs}.
\label{eq:E_kinetic}
\end{equation}
This amounts to less than a percent of the total hydrodynamic
energy carried by the supernova ejecta. 
The radiating shell contains a total electron energy of
\begin{equation}
E_{\rm e}\approx n\gamma m_ec^2\times {4\pi r^3\over \eta}= 2.0\times
10^{49}~\nu_8 \tau_{.6} A_3^4 r_{16}^{-6}~{\rm ergs},
\label{eq:E_e}
\end{equation}
and a magnetic energy of
\begin{equation}
E_{\rm B}\approx {B^2\over 8\pi} \times {4\pi r^3\over \eta}= 2.5\times
10^{43}\nu_8^{2} A_3^{-4} r_{16}^{11}~{\rm ergs}.
\label{eq:E_B}
\end{equation}

Our model predicts that the electrons and ions are nearly in {\it
equipartition},
\begin{equation}
{E_{\rm e}\over E_{\rm shock}}\approx 1.2 {\mu_2} A_3
V_{.3}^{-2}r_{16}^{-2}.
\label{eq:E_eE_B}
\end{equation}
This is consistent with our model assumption that the radio emission is
produced by thermal electrons. The post-shock magnetic field is several
orders of magnitude weaker than its equipartition amplitude.

Let us now consider the constraints imposed by the X-ray observations of
BeppoSAX. The radio radiation energy inside the shell volume is
\begin{equation}
E_{\rm radio}\approx \left({4L_{\rm radio}\over
c}\right)\left({r\over\eta}\right) =5.3 \times 10^{43} r_{16}~{\rm ergs},
\label{eq:E_rad}
\end{equation} 
where the factor of 4 is due to the fact that synchrotron radiation is
approximately isotropic in the shell.  Thus, the ratio between the
synchrotron luminosity and the inverse Compton luminosity due to scattering
of radio photons is
\begin{equation}
{L_{\rm IC-radio}\over L_{\rm syn}}={E_{\rm radio}\over E_{\rm B}}\approx 2.1
\nu_8^{-2} A_3^{4} r_{16}^{-10}.
\label{eq:IC_syn}
\end{equation}
A single Compton scattering boosts the synchrotron photon frequency up to a
value of $\sim (4/3) \gamma^2 \nu_c \approx 1.2\times 10^{15} \nu_8 A_3^2
r_{16}^{-4}~{\rm Hz}$, i.e. into the optical regime. The resulting optical
luminosity due to the upscattered synchrotron radiation is well below the
thermal supernova emission in this band -- which is $\sim4$ orders of
magnitude larger than the radio luminosity. The upper limit on the X-ray
flux of SN1998bw in the 1--10 keV band (Pian et al. 1998a,b) is obviously
in accordance with the negligible X-ray flux due to the scattering of radio
photons.

Inverse-Compton scattering of optical supernova photons by the relativistic
electrons yields a $\gamma$-ray flux which peaks at a photon energy of
$\sim 300$~keV and has a low-energy tail extending into the soft X-ray
range. For input radiation of flux $F_0$ at a frequency $\nu_0$, the
low-energy tail of the inverse-Compton emission due to electrons with a
Lorentz factor $\gamma$ is given by (Rybicki \& Lightman 1979), $f_\nu=
0.75\tau_T(\nu/\gamma^2\nu_0)F_0/\nu_0$, where $\tau_T$ is the Thomson
optical depth of the shell. For the supernova optical emission we
approximate $F_0/\nu_0$ by the peak flux of $\sim 10$~mJy in the V-band
($\nu_0=5\times10^{14}$Hz) and obtain
\begin{equation}
F_{\rm X}\approx10^{-13}\nu_8 \tau_{.6} A_3 r_{16}^{-2}~{\rm
erg~cm^{-2}~s^{-1}},
\label{eq:IC_SN}
\end{equation}
for the 1--10keV flux. The predicted flux is consistent with the BeppoSAX
upper limit, $\sim10^{-13}{\rm erg~cm^{-2}~s^{-1}}$, for
$r_{16}\ga1$. (Note that the BeppoSAX limit applies to day 8, when both the
optical and the radio fluxes are smaller by a factor $\sim2$ compared to
the corresponding fluxes at day 12 considered here).  Inverse-Compton
scattering of far infrared photons with initial wavelength $\sim30\mu$m,
produces a flux that peaks in the X-ray band. However, since the supernova
flux at low frequencies is roughly thermal, $f_\nu\propto\nu^2$, the
low-frequency tail of the inverse-Compton scattering of optical photons
dominates the X-ray flux.

Finally, we calculate the electron cooling time.  The synchrotron cooling
time is given by,
\begin{equation}
t_{\rm syn}\equiv {6\pi m_e c\over\gamma \sigma_{\rm T} B^2}
\approx 2\times10^2
\nu_8^{-2} A_3^3 r_{16}^{-6}~{\rm yrs},
\label{eq:t_syn}
\end{equation}
where $\sigma_{\rm T}$ is the Thomson cross-section. The
inverse-Compton cooling time equals,
\begin{equation}
t_{\rm IC}=\left({E_{\rm B}\over E_{\rm opt}}\right) \times t_{\rm Syn}
\approx 5 \nu_8^2 A_{3}^{-4} r_{16}^{10}~{\rm days},
\label{eq:t_IC}
\end{equation}
where, $E_{\rm opt}=L_{\rm opt} r/\eta c$, is the optical radiation energy
inside the shell volume, and we have used $L_{\rm opt}=10^{43}{\rm
erg~s^{-1}}$.  Inverse-Compton cooling might therefore be responsible for
the decline of the radio lightcurve immediately following the first peak.

\subsection{Observations After the Peak of the Radio Lightcurve}

Figure 2 shows the observational data around the second, weaker, radio peak
(33 days after GRB980425) and at the last observational epoch (58 days
after GRB980425). At both epochs, the spectrum is well fitted by synchrotron
emission of mono-energetic electrons, with a very small self-absorption
optical depth, $\tau_c=0.06$ and $\tau_c=0$ at the earlier and later times,
respectively. Since $\tau_c\ll 1$, the spectrum is essentially fitted by an
unabsorbed synchrotron model and for a given $r$ the observations provide
only two constraints on the three model parameters $n$, $\gamma$ and
$B$. Thus, the late time observations can not determine the model
parameters in the same way as observations near the first peak do (see
\S3.1). The decline in the self-absorption optical depth with time may be
due to a decrease in the ambient gas density with radius, as expected for a
wind.  Similarly, the complicated lightcurve structure might reflect
clumpiness in the spatial distribution of the ambient gas.

\section{Discussion}

The emission spectrum predicted by our model for mono-energetic electrons
provides an excellent fit to the observational radio data of SN1998bw (see
Figs. 1 and 2). This agreement implies that most of the radiating electrons
share the same energy, as expected if their distribution is thermal rather
than power-law in energy. Comparison of model predictions to data around
the peak of the radio lightcurve suggests that the total energy content of
the radiating electrons is comparable to the kinetic energy carried by the
ions in the forward shock, implying that the two species are in
equipartition at a temperature of $\sim 60$ MeV.  On the other hand, the
inferred magnetic field is much weaker than its equipartition amplitude and
could have originated from a frozen-in seed field that evolved
adiabatically with the ambient gas and did not suffer any dynamo
amplification behind the shock.  The field is likely to have been
carried-out with the wind that emerged from the progenitor's envelope prior
to the supernova. In fact, its predicted amplitude of $\sim 20$mG
[Eq.~(\ref{eq:magnetic_field})] is only an order of magnitude higher than
that expected from the adiabatic compression of the interstellar $\sim
\mu$G field as the interstellar medium density is increased by $\sim 5$
orders of magnitude to the value of the ambient gas density
[Eq.~(\ref{eq:density})].

The source radius required in order that our model be consistent with X-ray
observations near the peak of the radio emission, $\sim 12$ days after
GRB980425, is only $\sim 10^{16}~{\rm cm}$.  This implies a shock velocity
of $\sim 0.3c$ and a corresponding Lorentz factor of only 1.05. For these
parameters, the mass and energy associated with the radio emitting plasma
are $\sim 10^{-4}M_\odot$ and $\sim 10^{49}$~ergs, respectively. Our model
is therefore consistent with the modeling of the optical emission by
Woosley et al. (1998) who suggested that the outer $10^{-3}M_\odot$ of
the ejecta in SN1998bw might have moved at a speed of $\sim c/3$ and
carried $\sim 10^{50}$~ergs.

Based on these findings we conclude that the radio luminosity of 1998bw
could have been produced by the sub-relativistic shock in front of the
supernova ejecta. In our model, there is no need to invoke an
ultra-relativistic shock that might have been related to GRB980425. This
raises the possibility that GRB980425 was associated instead with the
transient X-ray source 1SAXJ1935.3-5252.

The large scatter in the radio luminosities of Type Ib/Ic supernovae (for
examples, see Weiler \& Sramek 1988; van Dyk et al. 1993; and Weiler et al.
1998) is not surprising in view of the fact that the total energy and mass
of the radio-emitting plasma are negligible fractions of the overall energy
and mass budgets of the supernova ejecta. The scatter might also be
enhanced by viewing angle variations due to non-spherical explosion
geometries (H\"oflich, Wheeler, \& Wang 1998).

Our discussion ignored various effects, such as the increase in the
self-absorption optical depth towards the limb of the source, the free-free
absorption in the surrounding stellar wind, or the nonzero time delay due
to the finite light crossing time across the source. Future work is
necessary in order to refine the detailed predictions of this model.

\acknowledgements This work was supported in part by the NASA grants
NAG5-3085 and NAG5-7039 (for A.L.).  A.L. thanks the Weizmann Institute for
its kind hospitality during the course of this work.

\clearpage
\newpage
\begin{figure}[b]
\vspace{2.6cm} \includegraphics{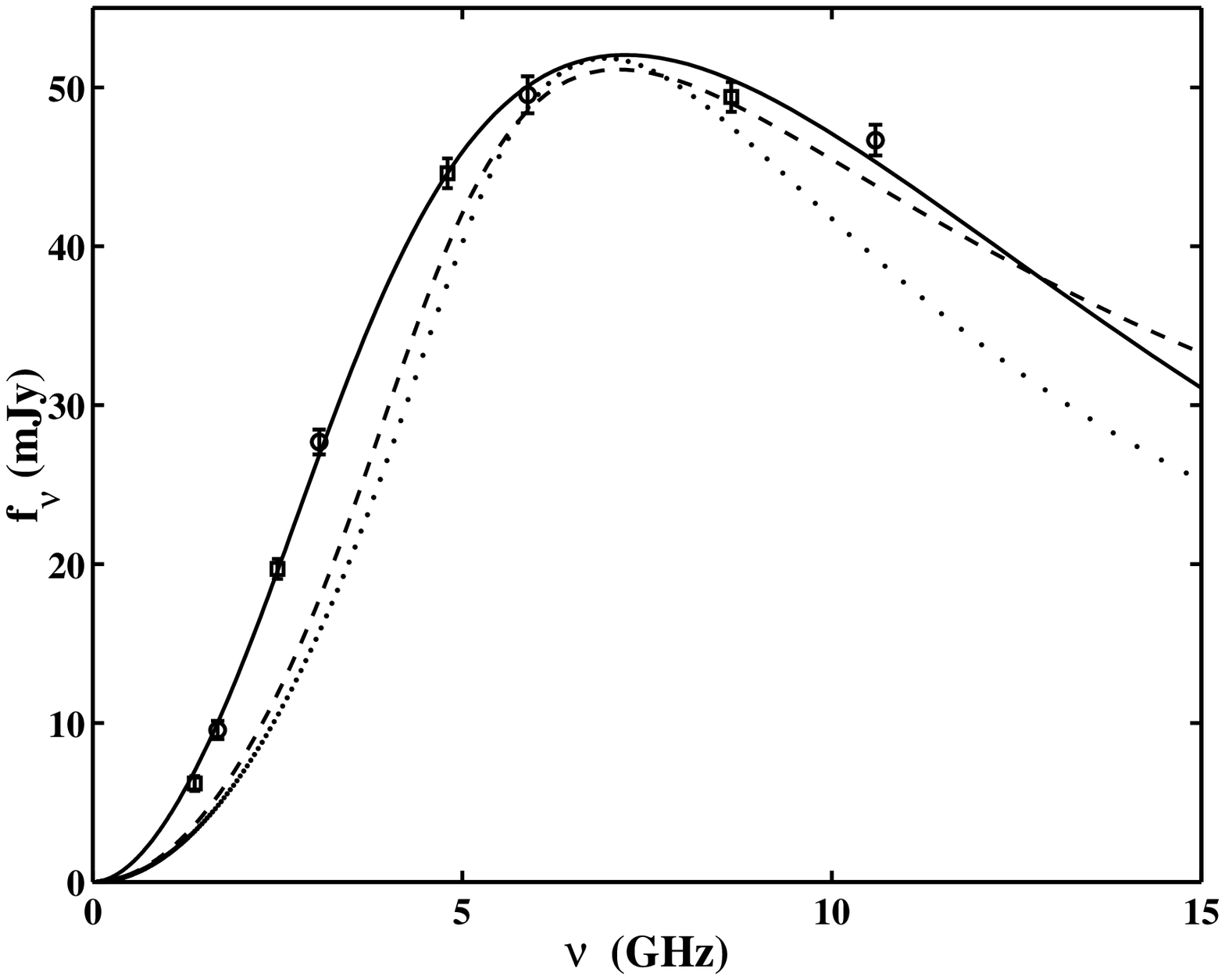}
\vspace*{4.5in}
\caption[Spectrum1] {\label{fig1:sources} Emission spectrum from SN1998bw
around the peak of the radio lightcurve. The data points from Kulkarni et
al. (1998) are shown 12 days (squares) and 15 days (circles) after
GRB980425 and are compared to theoretical predictions for a population of
electrons which is either mono-energetic (solid line), or power-law in
energy with a spectral index of $p=2$ (dashed) and $p=3$ (dotted).  The
frequency and flux of the second epoch observations (15 days after
GRB980425) are renormalized so as to show that the spectra at both
epochs have a similar shape (frequency multiplied by 1.23, flux by
1.24). The model parameters for mono-energetic electrons are $\tau_c=0.56$,
$A=3.3\times10^{-44}$~g, $\nu_c=7.6$~GHz [see Eq.~(\ref{eq:totalflux})].}
\label{fig:1} 
\end{figure}

\newpage
\begin{figure}[b]
\vspace{2.6cm}
\includegraphics{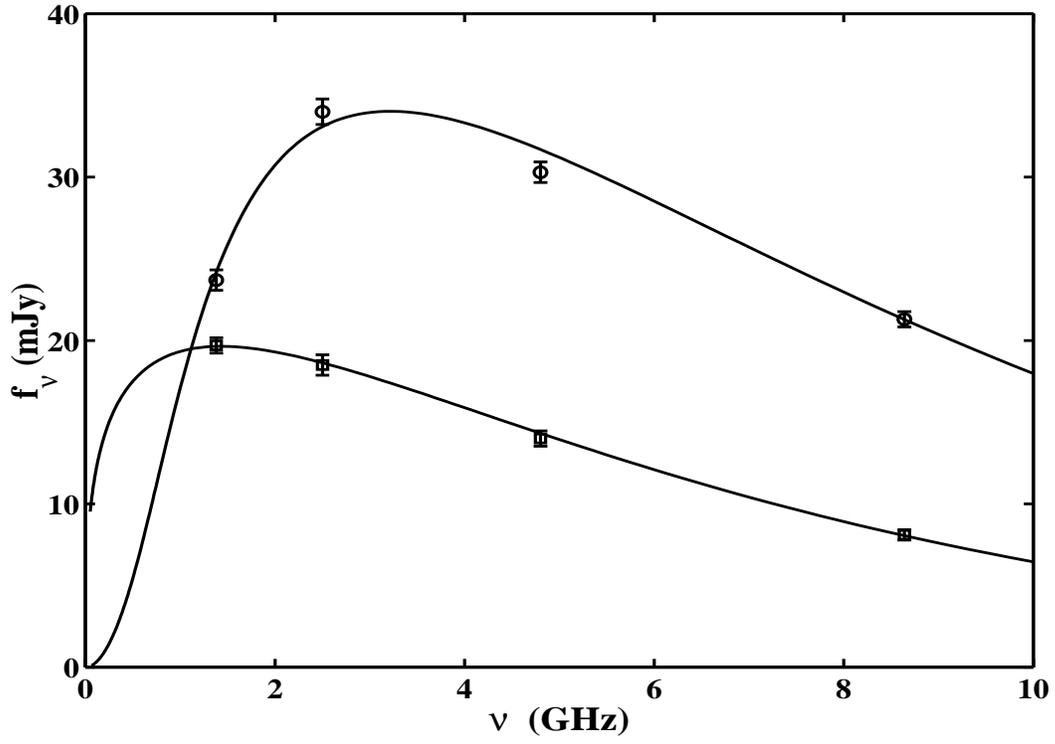}
\vspace*{4.5in}
\caption[Spectrum2] {\label{fig2:sources} Emission spectrum from SN1998bw
around the second radio peak (circles) and at the last observational epoch
(squares), 33 and 58 days after GRB980425 respectively (Kulkarni et
al. 1998). The data points are compared to the theoretical predictions for
a mono-energetic electron distribution with $\tau_c=0.06$ at the first
epoch and $\tau_c=0$ at the second epoch.}
\label{fig:2} 
\end{figure}

\end{document}